\documentclass[aps,prb,showpacs,twocolumn]{revtex4}
\usepackage{mathrsfs}
\usepackage{amssymb}
\usepackage{amsmath}
\usepackage{bm}
\usepackage{graphics}
\usepackage{natbib}

\begin{document}

\title{Fano-Josephson effect of Majorana bound states}
\author{Zhen Gao$^1$}
\author{Shu-Feng Zhang$^2$}
\author{Ying Zhao$^1$}
\author{Guangyu Yi$^1$}
\author{Wei-Jiang Gong$^1$}\email{gwj@mail.neu.edu.cn}

\affiliation{1. College of Sciences, Northeastern University, Shenyang
110819, China\\
2. Institute of Physics, Chinese Academy of Sciences, Beijing 100080, China}
\date{\today}

\begin{abstract}
We investigate the Josephson current in a Fano-Josephson junction formed by the direct coupling between two topological superconducting wires and their indirect coupling via a quantum dot. It is found that when two Majorana zero modes respectively appear in the wires, the Fano interference causes abundant Josephson phase transition processes. What is notable is that in the presence of appropriate direct and indirect inter-wire couplings, the fractional Josephson effect disappears and then such a structure transforms into a $0$-phase normal Josephson junction. On the other hand, if finite coupling occurs between the Majorana bound states at the ends of each wire, the normal Josepshon current is robustly in the $0$ phase, weakly dependent on the Fano effect. We believe that the results in this work are helpful for describing the Fano-modified Josephson effect.
\end{abstract}
\maketitle
\section{Introduction}
The energy of a tunnel junction between two superconductors (a Josephson junction) depends on the phase difference $\phi$ of the order parameter on the two sides of the junction. Its derivative vs $\phi$ exactly reflects the Josephson current flowing through the junction in the absence of an applied voltage. It is well known that in the Josephson junction formed by the coupling between two s-wave superconductors, the Josephson current is characterized by the formula $I_J \sim \sin\phi$ with its period $2\pi$.\cite{Golubov} With the development of low-dimensional semiconductor technique, one quantum dot (QD) and coupled-QD molecules can be fabricated to embed in the Josephson junction.\cite{QDJosephson,Tarucha,Meden} Such systems have accordingly attracted extensive investigation and the unique properties of QDs have been found to induce abundant Josephson phase transitions. As reported in the previous works, the QD-embedded Josephson junction shows $0$, $0'$, $\pi'$, and $\pi$ junction behaviors, respectively, with the enhancement of electron interaction.\cite{QD1,QD2,QD3,QD4} When a QD molecule is embedded in the Josephson junction, the quantum interference can cause the $0$-$\pi$ phase transition in an alternate way.\cite{AB1,AB2} Moreover, the Fano-Kondo effect in the QD molecule is able to induce the appearance of a bistable phase in the $0$-$\pi$ phase-transition process.\cite{Yigy}
\par
In recent years, topological superconductor (TS) has received considerable experimental and theoretical attention because Majorana zero-energy modes appear at the ends of the one-dimensional TS which can potentially be used for decoherence-free quantum computation.\cite{Majorana1,RMP1,Zhang1,Zhang2} In comparison with the conventional superconductor, the TS system shows new and interesting properties.\cite{Majorana2} For instance, in the proximity-coupled semiconductor-TS devices, the Majorana zero modes induce the zero-bias anomaly.\cite{Fuliang} A more compelling TS signature is the unusual Josephson current-phase relation. Namely, when the normal s-wave superconductor nano-wire is replaced by a TS wire with the Majorana zero modes, the current-phase relation will be modified to be $I_J \sim \sin{\phi\over2}$ and the period of the Josephson current vs $\phi$ will be $4\pi$. This is the so-called the fractional Josephson effect.\cite{Josephson1,Josephson2,Josephson3} Such a result can be understood in terms of fermion parity (FP). If the FP is preserved, there will be a protected crossing of the Majorana bound states (MBSs) at $\phi=\pi$ with perfect population inversion. As a result, the system cannot remain in the ground state as $\phi$ evolves from $0$ to $2\pi$ adiabatically.\cite{Aguado}
\par
In view of the previous results, one can understand that the FP is a nontrivial factor to regulate the fractional Josephson effect in the TS junction. It is known that QDs are able to accommodate electrons and the electron occupation number in QDs can be changed via shifting the QD levels. Consequently, when a QD molecule is introduced in the TS junction, the FP can be re-regulated and the fractional Josephson current can accordingly be modified. Moreover, some special QD geometries can induce the typical quantum interference mechanisms, e.g., the Fano interference,\cite{Fanormp} which are certain to play an important role in adjusting the fractional Josephson effect. As a result, interesting phase transitions can be anticipated in the QD-existed TS junction.
\par
In this work, we design a Fano-Josephson junction which is formed by the direct coupling between the two TS wires and their indirect coupling via a QD. We would like to carry out a comprehensive analysis about the influence of the Fano interference on the fractional Josepshon effect. As a result, we find that when Majorana zero-energy modes respectively appear in the TS wires, the Fano interference assists to drive abundant Josephson phase transition processes for conserving the FP. Moreover, in the presence of appropriate structural parameters, such a structure will change to be a normal Josephson junction with the $0$-phase Josephson current. On the other hand, if finite coupling occurs between the Majorana bound states in each wire, only the normal Josepshon effect will occur with the trivial role of the Fano interference.
\section{Theory}
\begin{figure}
\centering\scalebox{0.5}{\includegraphics{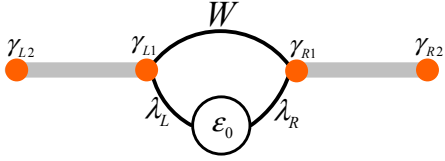}}
\caption{Schematic of a Fano-Josephson junction of Majorana bound states.} \label{Struct}
\end{figure}
The Josephson junction that we consider is illustrated in Fig.\ref{Struct}. In such a junction, the TS wires are described by two Kitaev chains.\cite{Kitaev} There are two kinds of couplings between the chains, i.e., the direct coupling and the indirect coupling via a QD. The Hamiltonian of this system can written as $H=\sum H_\alpha+H_D+H_T$. $H_\alpha$ is the Hamiltonian of the $\alpha$-th Kitaev chain, i.e., chain-$\alpha$. $H_D$ is the Hamiltonian of the embedded QD, and $H_T$ describes the inter-chain and the QD-chain couplings. They are respectively given by\cite{Rpp}
\begin{eqnarray}
H_\alpha&=&-\mu_\alpha\sum_{j=1}^{N_\alpha} c_{\alpha j}^\dag c_{\alpha j}-\sum_{j=1}^{N_\alpha-1}(t_\alpha c_{\alpha j}^\dag c_{\alpha, j+1}\notag\\
&&+|\Delta_\alpha| e^{i\theta_\alpha}c_{\alpha j} c_{\alpha, j+1}+h.c.),\notag\\
H_D&=&\varepsilon_0d^\dag d,\notag\\
H_T&=&-\lambda_L c^\dag_{LN}d-\lambda_R c^\dag_{R1}d-Wc^\dag_{LN}c_{R1}+h.c..
\end{eqnarray}
$c^\dag_{\alpha j}$ and $d^\dag$ ($c_{\alpha j}$ and $d$) are the operators to create (annihilate) an electron at the $j$-th site of chain-$\alpha$ and the QD. $\mu_\alpha$ is the onsite energy of the $j$-th site in chain-$\alpha$. $t_\alpha$ denotes the inter-site coupling, and the last term in $H_\alpha$ is the p-wave superconducting term. $\varepsilon_0$ is the QD level. In addition, $\lambda_\alpha$ denotes the coupling strength between the QD and chain-$\alpha$, and $W$ is the inter-chain coupling coefficient.
\par
The phase difference between the two Kitaev chains will drive finite Josephson current through them, which can be directly evaluated by the following formula
\begin{equation}
I_J= {2e\over \hbar}{\langle{\partial H}\rangle\over\partial \phi}
\end{equation}
where $\phi=\theta_R-\theta_L$ is the inter-chain phase difference and $\langle\cdots\rangle$ is the thermal average.
\section{Numerical results and discussions}
Based on the theory in the above section, we proceed to discuss the Josephson current through this structure. For simplicity, we focus on the case of zero temperature in the context.
\subsection{The case of Majorana zero modes}
\par
In the case of $N_\alpha=\infty$, the Josephson current occurs between two Majorana zero modes, one can therefore
project $H_\alpha$ onto the zero-energy subspace of $H_\alpha$ by sending $c_{LN}={1\over2}e^{-i\theta_L/2}\gamma_{L1}$ and $c_{R1}={i\over2}e^{-i\theta_R/2}\gamma_{R1}$ which yields an effective low-energy Hamiltonian\cite{Rpp}
\begin{eqnarray}
H_{\text{eff}}&=&-i{W\over2}\cos{\phi\over2}\gamma_{L1}\gamma_{R1}+\varepsilon_0\tilde{d}^\dag \tilde{d}-{\lambda_L\over2}\gamma_{L1}\tilde{d}\notag\\
&&-{i\lambda_R\over2}e^{i\phi/2}\gamma_{R1}\tilde{d}+h.c.,
\end{eqnarray}
with $\tilde{d}=e^{i\theta_L\over2}d$.
Here each MBS is the zero-energy superpositions of a
particle and a hole, thus the paired MBSs can be
fused into a Dirac fermion by defining $\gamma_{L1}=(f^\dag+f)$ and $\gamma_{R1}=i(f^\dag-f)$ ($f^\dag$ and $f$ are the fermionic creation and annihilation operators). As a result, $H_{\text{eff}}$ possesses its new form as
\begin{eqnarray}
H_{\text{eff}}&=&-W\cos{\phi\over2}(f^\dag f-\frac{1}{2})+\varepsilon_0 \tilde{d}^\dag \tilde{d}-{\lambda_L\over2}(f^\dag +f)\tilde{d}\notag\\
&&+{\lambda_R\over2}e^{i\phi/2}(f^\dag-f)\tilde{d}+h.c.
\end{eqnarray}
Since in such a system, only the FP is the good quantum number, we should discuss the Josephson current in the even- and odd-FP subspaces of the Fock space, respectively. With this idea, the basis $\{|00\rangle,|10\rangle,|01\rangle,|11\rangle\}$ should be regrouped according to the parity ($|n_f n_d\rangle=|n_f\rangle|n_d\rangle$, where $n_f=f^\dag f$ and $n_d=\tilde{d}^\dag \tilde{d}$). It is easy to find that the even-FP basis is $\{|00\rangle,|11\rangle\}$ which can be labeled by ${\cal P}=+1$, whereas $\{|10\rangle,|01\rangle\}$ is the odd-FP basis with ${\cal P}=-1$. Obviously, the presence of QD modifies the original FP of the TS junction. As a result, $H$ reduces to two $2\times 2$ matrixes according to the FP. First, in the case of even FP, the Fock state can be given by $|e\rangle=a|00\rangle+b|11\rangle$ and the matrix form of $H^{(e)}$ is expressed as
\begin{eqnarray}
H^{(e)}=
\left[
\begin{array}{cc}
 {W\over2} \cos{\phi\over2} & {\lambda_L\over2}+{\lambda_R\over2} e^{i\phi /2} \\
 {\lambda_L\over2}+{\lambda_R\over2} e^{-i\phi /2} & \varepsilon _0-{W\over2}\cos{\phi\over2}
\end{array}
\right].
\end{eqnarray}
On the other hand, in the case of odd FP, the Fock state is $|o\rangle=a|10\rangle+b|01\rangle$ and the matrix form of $H^{(o)}$ can be written as
\begin{eqnarray}
H^{(o)}=
\left[
\begin{array}{cc}
-{W\over2}\cos{\phi\over2} & -{\lambda_L\over2}+{\lambda_R\over2} e^{i\phi/2} \\
 -{\lambda_L\over2}+{\lambda_R\over2} e^{-i\phi/2} &\varepsilon _0+{W\over2}\cos{\phi\over2}
\end{array}
\right].
\end{eqnarray}
\par
Based on the above discussion, the Josephosen current of such a structure can be calculated with the help of the following formula $I_J({\cal P})={2e\over \hbar}{\langle{\partial H_{\text{eff}}({\cal P})}\rangle\over\partial \phi}= {2e\over \hbar}{\partial E_-({\cal P})\over\partial\phi}$. Via a straightforward calculation, we get the analytical expressions of $E_{-}({\cal P})$ and $I_J({\cal P})$. They are given by
\begin{eqnarray}
&&E_{-}({\cal P})={1\over2}\varepsilon_0-{1\over2}{\sqrt{(\varepsilon_0-{\cal P}W\cos{\phi\over2})^2+\Gamma({\cal P})}},\label{En}\\
&&I_J({\cal P})={e\over 4\hbar}{2{\cal P}(\lambda_L\lambda_R-W\varepsilon_0)\sin{\phi\over2}+W^2\sin\phi\over\sqrt{(\varepsilon_0-{\cal P}W\cos{\phi\over2})^2+\Gamma({\cal P})}}. \label{Jose}
\end{eqnarray}
with $\Gamma({\cal P})=\lambda^2_L+\lambda^2_R+2{\cal P}\lambda_L\lambda_R\cos{\phi\over2}$.
The result in Eq.(\ref{Jose}) shows that although the complicated geometry, the Josephson currents in different parities obey the relationship of $I_J({\cal P},\phi)\equiv I_J({\cal P}',\phi\pm2\pi)$. In view of this phenomenon, we would like to focus on the even-FP case (i.e., ${\cal P}=+1$) to clarify the influence of Fano interference on the Josephson effect.
\begin{figure}
\centering\scalebox{0.4}{\includegraphics{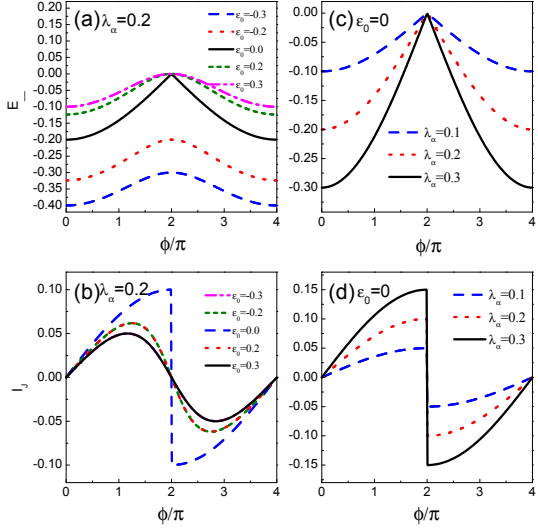}}
\caption{ The GS level and Josephson current in the case of $W=0$. (a)-(b) The influence of the QD level on $E_{-}$ and $I_J$ in the case of $\lambda_\alpha=0.2$. (c)-(d) The effects of the QD-chain couplings when $\varepsilon_0=0$. } \label{W00a}
\end{figure}
\par
To begin with, it is necessary to analyze the Josephson current in each channel. For the Josephson current contributed by the resonant channel, it can be discussed by supposing $W=0$. Surely, in such a case the expressions of the ground-state (GS) level and the Josephson current can be simplified. They are respectively written as
$E_{-}={1\over2}\varepsilon_0-{1\over2}{\sqrt{\varepsilon_0^2+\Gamma}}$ and
$I_J={e\over 2\hbar}\lambda_L\lambda_R\sin{\phi\over2}/\sqrt{\varepsilon_0^2+\Gamma}$.
Based on these results, we present the GS level and Josephson current influenced by the QD level and QD-chain couplings, as shown in Fig.\ref{W00a}. It clearly shows that $E_{-}$ and $I_J$ oscillate with $4\pi$ period, so the fractional Josepshon effect holds when one QD embeds in the TS junction formed by two Majorana zero modes. Since the minimum of $E_{-}$ always occurs at the point of $\phi=0$, the Josephson junction can be called the topological-$0$ junction (The concept \emph{topological-$0$} is introduced to describe the 0-phase current of the fractional Josephson effect). However, the QD level and QD-chain couplings play different roles in adjusting the properties of $E_{-}$ and $I_J$. In Fig.\ref{W00a}(a) where $\lambda_\alpha=0.2$, we find that with the increment of $\varepsilon_0$ to $\varepsilon_0=0$, the value of $E_{-}$ increases with the larger increase rate near the point of $\phi=2\pi$. At such a point, when the QD level tunes to be $\varepsilon_0=0$, the value of $E_{-}$ undergoes a sharp decrease. Next, further increasing $\varepsilon_0$ only induces a little increase of $E_{-}$ in the region away from the point of $\phi=2\pi$. Meanwhile, the change of $E_{-}$ becomes smooth again around this point. Next, in Fig.\ref{W00a}(b) we plot the Josephson current spectra affected by the shift of the QD level. It is seen that the Josephson current is a even function of $\varepsilon_0$, and that current amplitude is inversely proportional to the value of $|\varepsilon_0|$. In the case of $\varepsilon_0=0$, the Josephson current changes discontinuously at the point of $\phi=2\pi$. These are easy to understand with the help of the results in Fig.\ref{W00a}(a). Note, also, that in the case of $\varepsilon_0=0$, the good quantum coherence enhances the current amplitude. On the other hand, we see in Fig.\ref{W00a}(c)-(d) that the role of QD-chain couplings is relatively simple. Interpretively, the QD-chain couplings cannot change the oscillation manners of $E_{-}$ and $I_J$, whereas they only change the amplitudes of them.
\begin{figure}
\centering\scalebox{0.4}{\includegraphics{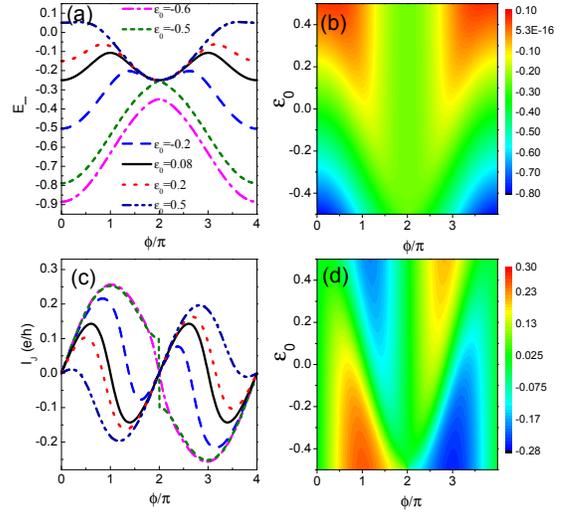}}
\caption{The GS level and Josephson current in the case of $W=0.5$ and $\lambda_\alpha=0.2$. (a)-(b) The influence of the QD level on $E_{-}$. (c)-(d) The effects of the QD level on the Josephson current.} \label{Fanoa}
\end{figure}
\par
Following the above results, we take $W=0.5$ to present the influence of the Fano interference on the Josephson current. For the nonresonant channel, $H_{\text{eff}}=-W\cos{\phi\over2}(n_f-\frac{1}{2})$ with $I_J={e\over\hbar}\sin{\phi\over2}(n_f-{1\over2})$, hence only the value of $n_f$ determines the FP and the Josephson current.\cite{Rpp} Next, when the two channels co-exist, $I_J(n_f=\pm1)$ will have the opportunity to simultaneously participate in the FP-conserved Fano interference which inevitably contributes to the Josephson effect. We present the numerical results in Fig.\ref{Fanoa} by taking $\lambda_{\alpha}=\lambda$. First, Fig.\ref{Fanoa} shows the spectra of $E_{-}$ and $I_J$ influenced by the change of the QD level. Here the QD-chain couplings are taken to be $\lambda=0.2$. In Fig.\ref{Fanoa}(a)-(b), we find that compared with the case of $W=0$, the QD level adjusts the curve of $E_{-}$ in a different way. Even from the case of $\varepsilon_0=-0.5$, the value of $E_{-}(\phi=2\pi)$ tends to be independent of the increase of $\varepsilon_0$. This can be understood via the formula of $E_{-}$. One can readily find that in the case of $\lambda_\alpha=\lambda$, $E_{-}={1\over2}(\varepsilon_0-|\varepsilon_0+W|)$. This means that when $\varepsilon_0+W\geq0$, $E_{-}$ is always equal to $-{1\over2}W$, independent of $\varepsilon_0$ when $\phi=2\pi$. Next in the other region, especially the region close to $\phi=0$ (or $4\pi)$, the increment of $\varepsilon_0$ efficiently induces the increase of $E_{-}$. It is easy to find from Eq.(\ref{En}) that at the point of $\phi=0$, the curve of $E_{-}$ begins to present a peak when the QD level increases to $\varepsilon_0=W$. As a result, in the case of $\varepsilon_0<-W$, the spectrum of $E_{-}$ shows up as a peak at the point of $\phi=2\pi$, whereas around this point a valley forms in the case $\varepsilon_0>W$. Note additionally that with the increase of $E_{-}(\phi=0)$, it has an opportunity to be equal to $E_{-}(\phi=2\pi)$. The condition for this result can be obtained, i.e., $\varepsilon_0W=\lambda^2$.
\par
The $\varepsilon_0$-adjusted change of $E_{-}$ directly leads to different oscillation behaviors of the Josephson current, as shown in Fig.\ref{Fanoa}(c)-(d). It shows that with the increase of $\varepsilon_0$ from $-0.6$ to $0.5$, the curve of $I_J$ vs $\phi$ changes from $I_J\sim\sin{\phi\over2}$ to $I_J\sim-\sin{\phi\over2}$. Meanwhile, the current amplitude varies nonlinearly. When the period of the Josephson current decreases to $2\pi$, its amplitude reaches the minimum. In addition, in the case of the QD level below the energy zero point, the current oscillation is relatively apparent due to the greater current amplitude. According to the previous works, the oscillation manner of the Josephson current is usually described by the Josephson current phase.\cite{Yigy} For the Josephson current in this figure, it can be considered to be in the topological-$0$ phase in the case of $\varepsilon_0<-W$. In the region of $-W<\varepsilon_0<{\lambda^2\over W}$, the Josephson current can be considered to be in the topological-$0'$ phase, since the local minimum of $E_{-}$ emerges around the point of $\phi=2\pi$. Next, with the further increase of $\varepsilon_0$ to $\varepsilon_0=W$, the topological-$\pi'$ phase comes into being with a local minimum of $E_{-}$ around the point of $\phi=0$. Finally, at the limit of $\varepsilon_0>W$, only one valley appears around the point of $\phi=2\pi$ in a period, and then the Josephson current enters its topological-$\pi$ phase. Therefore, we readily find that the shift of the QD level gives rise to the abundant phase transition results. In addition, we would like to point out that the condition of $\varepsilon_0W=\lambda^2$ can also satisfy the result of $E_{-}(\pi+\phi)=E_{-}(3\pi+\phi)$ for any $\phi$. This exactly means that in such a case, the fractional Josephson effect vanishes but a normal Josephson effect occurs in the Junction with $2\pi$ period. Besides, since the global minimum of $E_{-}$ emerges at the point of $\phi=0$, the normal Josephson current is certainly in the $0$ phase.

\par
Up to now, we can conclude that in the TS junction of the Majorana zero modes, the Fano interference plays a nontrivial role in modifying the fractional Josephson effect in the energy region of $-W<\varepsilon_0<W$ where the fermion occupation in the QD is not fixed. The main results include the complicated phase transition behaviors as well as the halving of the current period. On the other hand, in the region of $|\varepsilon_0|>W$, the Fano interference becomes weak but only the nonresonant channel contributes to the Josephson current. Take the case of $\varepsilon_0>W$ as an example, one can understand that in such a case $n_d\sim0$, hence $I_J({\cal P}=+1)\sim-\sin{\phi\over2}$. In addition, it can be understood that in the odd-FP case, changing $\varepsilon_0$ will lead to the opposite phase transition process in such a TS junction due to the fact that $I_J({\cal P}, \phi)=I_J({\cal P}', \phi\pm2\pi)$. All these results are certain to be helpful for clarifying the Fano-modified fractional Josephson effect.
\subsection{The case of nonzero inter-MBS couplings}
\begin{figure}
\centering\scalebox{0.38}{\includegraphics{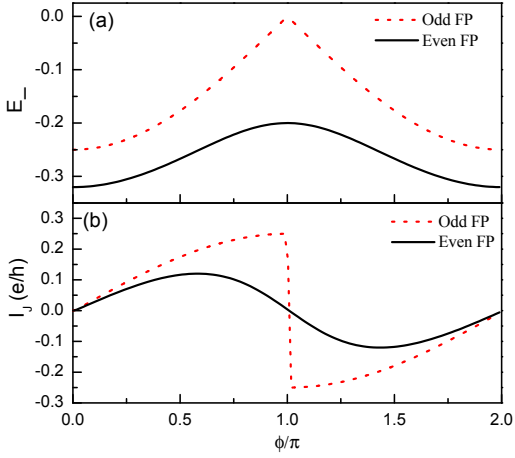}}
\caption{(a) The different-FP GS levels in the case of $\varepsilon_\alpha=0.1$ and $W=0.5$. (b) The corresponding Josephson currents.} \label{Em01b}
\end{figure}

In the case of finite coupling between the MBSs in each Kitaev chain, the effective Hamiltonian of our considered system can directly be written as
\begin{eqnarray}
H_{\text{eff}}&=&-i{W\over2}\cos{\phi\over2}\gamma_{L1}\gamma_{R1}+i\varepsilon_L\gamma_{L1}\gamma_{L2}
+i\varepsilon_R\gamma_{R1}\gamma_{R2}\notag\\&&+\varepsilon_0\tilde{d}^\dag \tilde{d}-{\lambda_L\over2}\gamma_{L1}\tilde{d}
-{i\lambda_R\over2}e^{i\phi/2}\gamma_{R1}\tilde{d}+h.c..
\end{eqnarray}
In this equation, $\varepsilon_\alpha$ denotes the coupling strength between the MBSs in chain-$\alpha$.
By defining $\gamma_{L1}=f^\dag_L+f_L$, $\gamma_{L2}=i(f^\dag_{L}-f_L)$, $\gamma_{R1}=i(f^\dag_R-f_R)$, and $\gamma_{R2}=f^\dag_{R}+f_R$, this Hamiltonian will transform into
\begin{eqnarray}
H_{\text{eff}}&=&{W\over2}\cos{\phi\over2}(f^\dag_Lf^\dag_R-f^\dag_Lf_R+f_Lf^\dag_R-f_Lf_R)+\varepsilon_0n_d\notag\\
&&+\varepsilon_L(2n_L-1)+\varepsilon_R(2n_R-1)-{\lambda_L\over2}(f_L^\dag+f_L)\tilde{d}\notag\\
&&+{\lambda_R\over2}e^{i\phi/2}(f^\dag_R-f_R)\tilde{d}+{\lambda_L\over2}\tilde{d}^\dag(f_L^\dag+f_L)\notag\\
&&+{\lambda_R\over2}e^{-i\phi/2}\tilde{d}^\dag(f_R-f^\dag_R),
\end{eqnarray}
where $n_L=f^\dag_Lf_L$, $n_R=f^\dag_Rf_R$, and $n_d=\tilde{d}^\dag \tilde{d}$.
Similar to the discussion manner in the above subsection, we would like to discuss the Josephson currents in the nonresonant and resonant channels, respectively, for presenting the Fano-modified Josephson effect. The property of the nonresonant channel can be clarified by taking $\lambda_\alpha=0$. Accordingly, the Fock state can be built on the basis of $\{|n_Ln_R\rangle\}$. In the even-FP case, the Fock state can be given by $|e\rangle=a|00\rangle+b|11\rangle$ and the corresponding matrix form of $H^{(e)}$ is
$H^{(e)}=\left[
\begin{array}{cc}
 -\varepsilon_L-\varepsilon_R & {W\over2} \cos{\phi\over2} \\
{W\over2} \cos{\phi\over2}& \varepsilon_L+\varepsilon_R
\end{array}
\right]$.
Via a simple derivation, the eigen-energies can be obtained, i.e., $E^{(e)}=\pm\sqrt{(\varepsilon_L+\varepsilon_R)^2+{W^2\over8}\cos\phi+{W^2\over8}}$. On the other hand, for the odd-FP case, $|o\rangle=a|01\rangle+b|10\rangle$, and
$H^{(o)}=\left[
\begin{array}{cc}
 -\varepsilon_L+\varepsilon_R & -{W\over2} \cos{\phi\over2}\\
-{W\over2} \cos{\phi\over2} &\varepsilon_L-\varepsilon_R
\end{array}
\right]$.
As a consequence, we get the result that $E^{(o)}=\pm\sqrt{(\varepsilon_L-\varepsilon_R)^2+{W^2\over8}\cos\phi+{W^2\over8}}$. These results clearly show that the GS levels in different fermion parities oscillate in phase with its $2\pi$ period. Accordingly, only the normal Josephson effect comes into being with its analytical expression
$I_J({\cal P})={e\over 8\hbar}W^2\sin\phi/\sqrt{(\varepsilon_L+{\cal P}\varepsilon_R)^2+{W^2\over8}\cos\phi+{W^2\over8}}$.
The appearance of the normal Josephson current can be understood as follows. The two MBSs at the ends of one TS wire allow for the hybridization of two states of the same FP. This results in residual splittings at $\phi=\pi$ which destroy the fractional effect as the system remains in the ground state for all $\phi$.\cite{Pikulin} In Fig.\ref{Em01b}, we take $W=0.5$ and $\varepsilon_\alpha=0.1$ and plot the spectra of the GS levels and the Josephson currents in different FPs. It can clearly be found that except the in-phase oscillation of the GS levels in different FPs, the odd-FP GS level presents its larger amplitude. In the case of $\phi=\pi$, $E^{(o)}_{-}$ has an opportunity to reach the Fermi level and then decreases sharply, which induces the discontinuous change of the Josephson current. Although the different Josephson currents in the two FPs, they are always located in the $0$ phase because of the global minima at the point of $\phi=0$.
\begin{figure}
\centering\scalebox{0.4}{\includegraphics{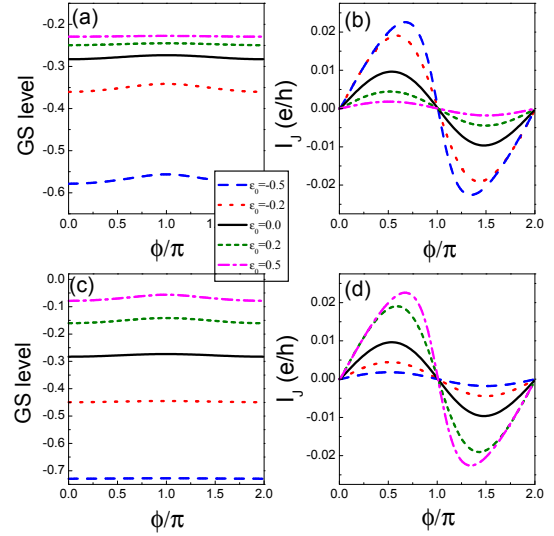}}
\caption{The GS level and Josephson current influenced by the change of QD level. The parameters are taken to be $W=0.0$, $\varepsilon_\alpha=0.1$, and $\lambda_\alpha=0.2$. (a)-(b) The GS level and Josephson current in the even-FP case. (c)-(d) The results in the odd-FP case.} \label{W00b}
\end{figure}
\par
The presence of QD inevitably re-regulates the FP of the TS junction. We next discuss the Josephson effect in the case of nonzero couplings, i.e., $\lambda_\alpha\neq0$. In such a case, the Fock state should be built on the basis $\{|n_Ln_Rn_d\rangle\}$. In the situation of even FP, the corresponding Fock state can be written as $|e\rangle=a_1|000\rangle+a_2|011\rangle+a_3|101\rangle+a_4|11 0\rangle$ with the matrix form of $H^{(e)}$
\begin{eqnarray}
&&H^{(e)}=\notag\\
&&\left[
\begin{array}{cccc}
 -\varepsilon_L-\varepsilon_R & {\lambda_R\over2} e^{i\phi /2} & {\lambda_L\over2} &{W\over2} \cos{\phi\over2} \\
 {\lambda_R\over2} e^{-i\phi /2} &-\varepsilon_L+\varepsilon_R+\varepsilon_0 & -{W\over2} \cos{\phi\over2}&{\lambda_L\over2}\\
{\lambda_L\over2} &-{W\over2} \cos{\phi\over2} &\varepsilon_L-\varepsilon_R+\varepsilon_0&{\lambda_R\over2} e^{-i\phi /2} \\
{W\over2} \cos{\phi\over2}& {\lambda_L\over2}& {\lambda_R\over2} e^{i\phi /2}&\varepsilon_L+\varepsilon_R
\end{array}
\right].\notag
\end{eqnarray}
Alternatively, for the odd-FP case, $|o\rangle=b_1|001\rangle+b_2|010\rangle+b_3|100\rangle+b_4|111\rangle$ and
\begin{eqnarray}
&&H^{(o)}=\notag\\
&&\left[
\begin{array}{cccc}
 -\varepsilon_L-\varepsilon_R+\varepsilon_0 & {\lambda_R\over2} e^{-i\phi /2} & -{\lambda_L\over2} &{W\over2} \cos{\phi\over2} \\
 {\lambda_R\over2} e^{i\phi /2} &-\varepsilon_L+\varepsilon_R & -{W\over2} \cos{\phi\over2}&-{\lambda_L\over2}\\
-{\lambda_L\over2} &-{W\over2} \cos{\phi\over2} &\varepsilon_L-\varepsilon_R&{\lambda_R\over2} e^{i\phi /2} \\
{W\over2} \cos{\phi\over2}& -{\lambda_L\over2}& {\lambda_R\over2} e^{-i\phi /2}&\varepsilon_L+\varepsilon_R+\varepsilon_0
\end{array}
\right].\notag
\end{eqnarray}
By taking $W=0$, we can discuss the Josephson current through the resonant channel. Since the analytical expressions of the GS level and Josephson current cannot be written out, we have to only present the numerical results, as shown in Fig.\ref{W00b}. The relevant parameters are taken to be $\varepsilon_\alpha=0.1$ and $\lambda_\alpha=0.2$. In this figure, we find that the current period also decreases to $2\pi$ and the FPs do not change the oscillation manner of the GS levels (so as to the Josephson currents). Besides, in such a case, only the 0-phase Josephson current can be observed with $I_J\sim\sin\phi$. The FP influence is manifested as follows. In the even-FP case, the increase of $\varepsilon_0$ can weaken the oscillation of the GS level and the Josephson current is suppressed gradually, as shown in Fig.\ref{W00b}(a)-(b). However, in the odd-FP case, increasing $\varepsilon_0$ strengthens the oscillation of the GS level, leading to the increase of the current amplitude [Fig.\ref{W00b}(c)-(d)]. In comparison with the currents in Fig.\ref{W00b}(b) and Fig.\ref{W00b}(d), we find that the Josephson currents in different FPs obey the relationship of $I_J({\cal P},\varepsilon_0)=I_J({\cal P}',-\varepsilon_0)$. Consequently, in the case of $\varepsilon_0=0$, the Josephson currents will be irrelevant to the parity of the fermion number. It is additionally notable that in the case of $\varepsilon_\alpha\neq0$, the current magnitude is much smaller than that in the case of Majorana zero modes (i.e., $\varepsilon_\alpha=0$).
\par
\begin{figure}
\centering\scalebox{0.4}{\includegraphics{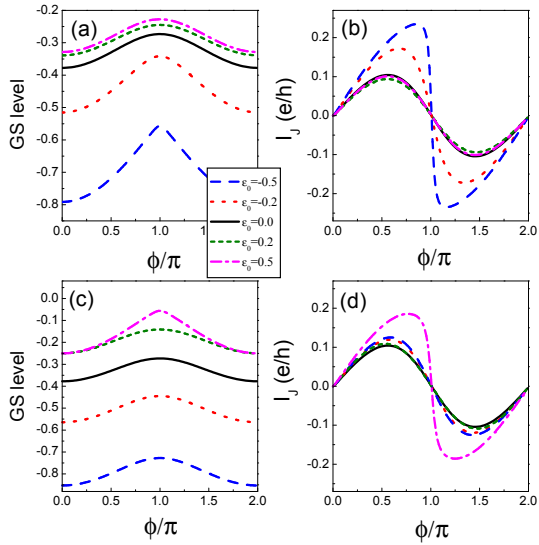}}
\caption{The GS level and Josephson current influenced by the Fano effect. The parameters are taken to be $W=0.5$, $\varepsilon_\alpha=0.1$, and $\lambda_\alpha=0.2$. (a)-(b) The even-FP results. (c)-(d) The results in the odd-FP case.} \label{Fanob}
\end{figure}
In the following, we expand the discussion about the Fano-modified Josephson effect. In Fig.\ref{Fanob}, we take $W=0.5$ and $\lambda_\alpha=0.1$ to calculate the GS levels and Joesphson currents in different FPs. For the inter-MBS coupling in each Kitaev chain, we also choose $\varepsilon_\alpha=0.1$. It can be clearly found that different from the case of Majorana zero modes, the influence of $\varepsilon_0$ on the GS levels and Josephson currents is just similar to the result of $W=0$ in respective FPs. To be concrete, in each FP, the global minimum of the GS level only occurs at the point of $\phi=0$ and the Josephson currents always keep to be in the $0$ phase regardless of the QD-level shift. Also, in the even-FP case, increasing $\varepsilon_0$ weakens the oscillation of the GS level and the Josephson current is suppressed gradually [See Fig.\ref{Fanob}(a)-(b)]; In the odd-FP case, the increase of $\varepsilon_0$ strengthens the oscillation of the GS level and the amplitude of $I_J$, as shown in Fig.\ref{Fanob}(c)-(d). On the other hand, the characteristics of this geometry can be observed. Firstly, $I_J({\cal P},\varepsilon_0)$ is not equal to $I_J({\cal P}',-\varepsilon_0)$. Secondly, the Josephson current tends to be independent of the QD-level shift in the even-FP case of $\varepsilon_0 >0$, and similar phenomenon occurs in the odd-FP case of $|\varepsilon_0|<0.2$. Moreover, in such two cases, the Josephson currents are almost the same. Surely, all these results should be attributed to the Fano interference. It should be emphasized that due to the robustness of the 0-phase currents contributed by the two channels, the Fano effect does not induce any Josephson phase transition here. In addition, we can note that in the even(odd)-FP case of $\varepsilon_0<-0.5$ ($\varepsilon_0>0.5$), the Fano effect becomes relatively weak, and the nonresonant channel contributes dominantly to the Josepshon current.

\section{summary}
In summary, we have investigated the Josephson current in a Fano-Josephson junction formed by the direct coupling between the two TS wires and their indirect coupling via a QD. As a result, it has been found that when two Majorana zero modes respectively appear in the TS wires, the Fano interference drives abundant Josephson phase transition processes for conserving the FPs. In the even-FP case, the shift of QD level induces the occurrence of the topological-$0$, topological-$0'$, topological-$\pi'$, and topological-$\pi$ phases. Moreover, in the presence of appropriate direct and indirect inter-wire couplings, the fractional Josephson effect disappears and this system is simplified to be a normal Josephson junction with the $0$-phase Josephson current. Next in the odd-FP case, the opposite result comes into being. Alternatively, if the inter-MBS coupling is nonzero in each TS wire, the Fano effect will make a weak contribution to the $0$-phase normal Josepshon current due to the FP-independence of the Josephson current. We believe that the results in this work are helpful for describing the Fano-modified Josephson effect.
\section*{Acknowledgments}
This work was financially supported by the Natural Science Foundation of Liaoning province of China (Grant No.
2013020030), the Liaoning BaiQianWan Talents Program (Grant No. 2012921078), and the Fundamental Research Funds for the Central Universities (Grant No. N130505001). W. J. G. thanks Fan Zhang for the helpful discussions.

\clearpage

\bigskip

\end{document}